# A Small Step for Epitaxy, a Large Step Towards Twist Angle Control in 2D Heterostructures

Oliver Maßmeyer[1]*, Jürgen Belz[1], Badrosadat Ojaghi Dogahe[1], Maximilian Widemann[1], Robin Günkel[1], Johannes Glowatzki[1], Max Bergmann[1], Sergej Pasko[2], Simonas Krotkus[2], Michael Heuken[2], Andreas Beyer[1] and Kerstin Volz[1]*

[1]Material Sciences Center and Department of Physics, Philipps-Universität Marburg, Germany

[2] AIXTRON SE, Herzogenrath, Germany

*Corresponding authors:
oliver.massmeyer@physik.uni-marburg.de
kerstin.volz@physik.uni-marburg.de

**ToC Figure:**

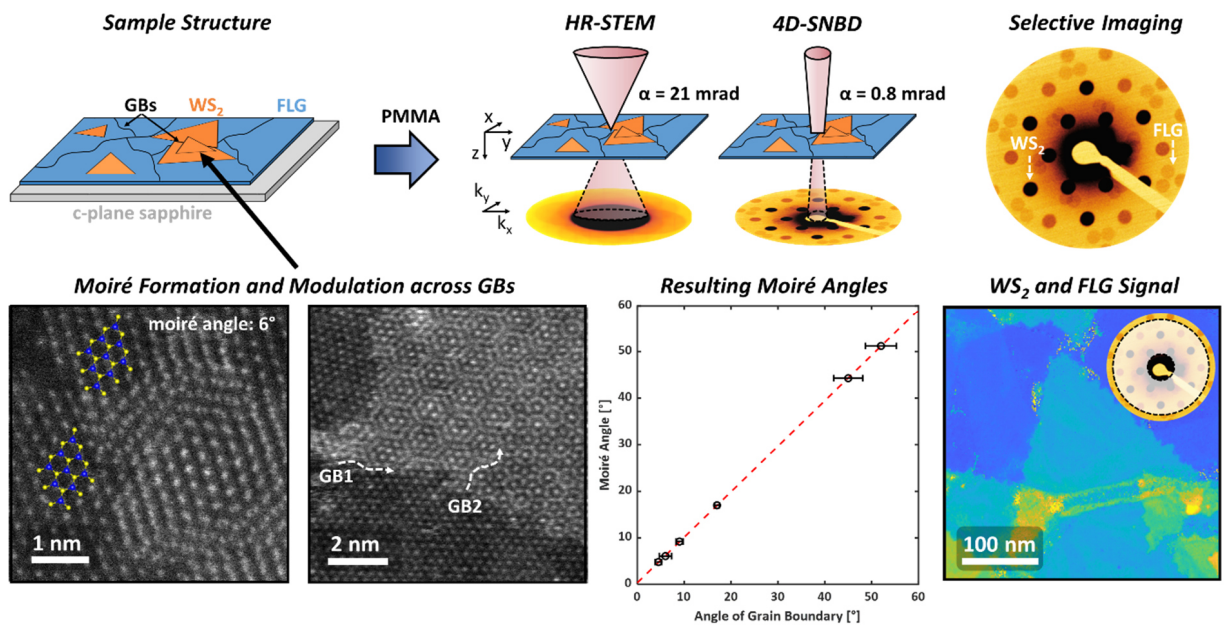

**Vertical 2D heterostructures of WS2 and few-layer graphene are studied by 4D scanning nanobeam diffraction**. Grain boundaries in the underlying 2D layer are shown to be essential for moiré formation in bottom-up metal-organic chemical vapor deposition synthesis. Controlling the angle of the grain boundary allows twist-angle-controlled growth of moiré structures.

## Abstract

**Two-dimensional (2D) materials have received a lot of interest over the past decade. Especially van der Waals (vdW) 2D materials, such as transition metal dichalcogenides (TMDCs), and their heterostructures exhibit semiconducting properties that make them highly suitable for novel device applications. Controllable mixing and matching of the 2D materials with different properties and a precise control of the in-plane twist angle in these heterostructures are essential to achieve superior properties and need to be established through large-scale device fabrication. To gain fundamental insight into the control of these twist angles, 2D heterostructures of tungsten disulfide (WS$_2$) and graphene grown by bottom-up synthesis via metal-organic chemical vapor deposition (MOCVD) are investigated using a scanning transmission electron microscope (STEM).**

Specifically, the combination of conventional high-resolution imaging with scanning nano-beam diffraction (SNBD) using advanced 4D STEM techniques is used to analyze moiré structures. The latter technique is used to reveal the epitaxial alignment within the WS$_2$/Gr heterostructure, showing a direct influence of the underlying graphene layers on the moiré formation in the subsequent WS$_2$ layers. In particular, the importance of grain boundaries within the underlying WS$_2$ and Gr layers for the formation of moiré patterns with rotation angles below 2° is discussed.

**Keywords: 2D, moiré, MOCVD, STEM, WS$_2$, graphene, grain boundaries**

1. Introduction

Ever since the first isolation of graphene (Gr) and the subsequent award of the Nobel Prize in 2010, two-dimensional (2D) materials have received increasing attention.[1–4] Especially heterostructures of van der Waals (vdW) 2D materials are widely discussed for novel device applications. Due to their geometric constraints, each 2D layer acts as both bulk material and interface, resulting in very different properties compared to 3D materials. In particular, the relatively weak vdW interaction removes the restriction of lattice constant matching between the 2D layer and the underlying substrate as well as between different 2D layers, allowing an almost arbitrary in-plane twist angle in 2D heterostructures.[5,6] In contrast to the weak bonding, the charge transfer between the individual 2D layers can be very large, leading to strong electric fields and thus interesting possibilities in band-structure engineering. This opens up unprecedented possibilities to combine 2D materials, e.g. with different electronic properties, leading to the possibility to study a wide range of fundamental physics, such as moiré patterns or interlayer excitons.[7–15]

Among the many 2D materials, special attention in research is given to the transition metal dichalcogenides (TMDCs), which exhibit a broad range of electronic properties allowing them to act as insulators, semiconductors or (semi-)metals.[2] A specific example is a vertical heterostructure of WS$_2$ and graphene, which has been discussed as a promising candidate for a field-effect tunneling transistor.[2,7] To realize these device structures, a precise control of the material composition, high uniformity as well as large-scale production is desired. Many preparation techniques are discussed in the literature to obtain these heterostructures in the desired quality.[3,16–21] On the one hand, top-down approaches such as artificial stacking by exfoliation and transfer techniques including polymers (e.g. polymethylmethacrylat, PMMA),[21] elastomer stamps (e.g. polydimethylsiloxane, PDMS)[20] or thermal release tapes (TRTs)[17] are used. While low cost and readily available, such vertical stacking methods suffer from the inability to produce high homogeneity on a large scale. On the other hand, bottom-up approaches for 2D growth, such as molecular beam epitaxy (MBE)[22–28] or chemical vapor deposition (CVD)[29–36] are currently explored. Especially MBE and metal-organic chemical vapor deposition (MOCVD)[35–38] are well-established techniques used for mass production in the semiconductor industry and allow the wafer-scale growth of materials with high crystalline quality.

However, there is still a lack of proper control over the stacking orientation and twist angle alignment during the growth of 2D homo- and heterostructures.[39–41] To gain a better understanding and a possible way to control the stacking and the twist angle, in this study, vertical 2D heterostructures of WS$_2$ on few-layer graphene (FLG) grown on c-plane sapphire substrates by MOCVD are investigated. We combine 4D scanning nano-beam diffraction (4D-SNBD) data obtained in a high-resolution scanning transmission electron microscope

(STEM) with atomic force microscopy (AFM) and scanning electron microscopy (SEM) to analyze the orientation relationship within the 2D heterostructure as well as to the underlying sapphire substrate. The 4D-SNBD data is subsequently used to disentangle the signals from the $WS_2$ and FLG layers, revealing moiré structures and grain boundaries (GB) in the FLG. In addition, the 4D-SNBD data highlights the critical role of GBs and the orientation of the underlying $WS_2$ and FLG layers in the orientation and formation of moiré patterns in the overlying $WS_2$. This finding potentially opens up a way to control the twist angle in these heterostructures during fabrication.

## 2. Methods

All 2D homo- and heterostructures are grown by MOCVD in a 19x2 inch Close Coupled Showerhead AIXTRON MOCVD reactor on c-plane sapphire substrates with an off-cut of 0.2° towards m-plane. To fabricate the $WS_2$/graphene heterostructure, graphene was first grown on sapphire using methane as a precursor at 1400 °C with a surface pretreatment step in $H_2$ as described in Ref.[38] Subsequent $WS_2$ nucleation was achieved on the graphene/sapphire template using tungsten hexacarbonyl ($W(CO)_6$) and ditertiarybutylsulfide (DTBS) as precursors at a growth temperature of 700 °C with a growth time of 1800 s.[42,43]

In a first step, the grown structures are characterized in a *Digital Instruments* IIIa AFM to obtain information about the sample topography. The images are analyzed and processed using *Gwyddion* software. In addition, a JEOL JIB-4601F SEM is used to image the sample on a larger scale and to analyze the orientation of the grown $WS_2$ triangles with respect to the c-plane sapphire. For the latter, multiple regions of the wafer with triangles are imaged at high contrast settings to allow for proper edge detection. This is then used to determine the distribution of the rotation angle of the triangles with respect to the a-plane (11-20) of the sapphire substrate (wafer edge).

The 2D heterostructures are then transferred to lacey carbon covered copper grids for (scanning) transmission electron microscopy ((S)TEM) by an etchant-free transfer method using PMMA.[21] In a first step the PMMA is spin-coated onto the 2D heterostructure on sapphire, followed by immersion of the sample in hot water at 80 °C. Due to the difference in hydrophilicity between the substrate and the PMMA-coated 2D material, water intercalates between the two, causing them to separate and the 2D film to float on the water surface. The separated 2D film is then picked up with the TEM grid and baked at 100°C on a hot plate to achieve better adhesion of the film to the grid. Finally, the PMMA is dissolved with dichloromethane and rinsed with isopropanol (IPA) and deionized water to remove any residual solvent. In addition, the samples are baked under vacuum at approximately 130 °C prior to STEM analysis.

The transferred samples are then characterized in a double aberration corrected JEOL 2200FS STEM operating at 80 kV and 200 kV. For high resolution imaging, a semi-convergence angle of 21 mrad and an inner collection angle in the range of 40 to 80 mrad and an outer collection angle of 160 to 280 mrad are used. This results in typical annular dark field (ADF) imaging conditions.[44,45] For diffractive imaging, the convergence angle is reduced to 0.8 mrad to avoid the complex diffraction pattern of overlapping diffraction discs typical of high convergence angles.[46] Due to the difference in lattice constant between $WS_2$ and Gr, the signal from $WS_2$ can be distinguished from the underlying FLG. For each scan position (x,y), the corresponding diffraction pattern (kx,ky) is detected by a fast pixelated pn-CCD detector[47], resulting in a 4D data set. In order to utilize the full-well capacity of the camera, which is the maximum charge a single pixel can hold before saturation, to obtain a higher signal for both the FLG and the $WS_2$, we neglected the anti-blooming mode and accounted for the blooming of the direct beam by moving the direct disc out of the detector. Due to the symmetry of the diffraction pattern, this approach allowed us to image a much larger diffraction space while avoiding oversaturation of the brighter diffraction spots. In addition, averaged diffraction patterns are recorded with a Gatan Ultrascan camera.

The 4D-SNBD data sets are evaluated by a combination of MATLAB programs. First, the data sets are post-processed as described in detail in Supporting Information S1.
By processing the 4D data sets, different virtual images can be generated, which are used to separate the signals from $WS_2$ and FLG to allow orientation mapping of each layer in the vertical 2D heterostructures. For orientation mapping, the intensity obtained within the hexagonal masks positioned around the diffraction spots is evaluated as a function of rotation angle and matched with the material symmetry.

## 3. Results and Discussion

We first discuss the main features of the sample including the orientation of the $WS_2$ triangles with respect to the a-plane (11-20) of the sapphire substrate, followed by an in-depth investigation and correlation of the influence of the FLG orientation on that of the $WS_2$. Based on this analysis, we show that moiré structures in the $WS_2$ are introduced by underlying GBs of $WS_2$ and graphene. This finding suggests a promising approach for epitaxial growth of specific moiré structures by controlling the structure of the GBs.

### 3.1. Vertical 2D Heterostructure of $WS_2$/FLG on Sapphire

An AFM image of the vertical 2D heterostructure of $WS_2$/FLG on sapphire showing the topography of the sample is shown in Figure 1 a). The sample exhibits homogeneous regions of $WS_2$ triangles in between chains of $WS_2$ triangles, which are caused by nucleation at wrinkles in the underlying FLG. These wrinkles are caused by the difference in thermal expansion between the sapphire and the FLG as it cools from the growth temperature of the FLG to room temperature. The $WS_2$ is then grown on the FLG/sapphire in a separate run. Within the homogeneous regions of the sample, a high density of $WS_2$ triangles is observed on top of the coalesced FLG layer. The $WS_2$ triangles are mostly separated and only coalesce at favorable positions on the sample. In addition, nucleation of additional $WS_2$ triangles is usually found as a $2^{nd}$, $3^{rd}$ or $4^{th}$ layer stacked on top of the $1^{st}$ $WS_2$ triangles.
Since the $WS_2$/FLG heterostructure detaches from the substrate during the PMMA transfer, information about the original relationship to the sapphire substrate can only be obtained by analyzing the structures prior to the STEM investigations. Therefore, we determine the orientation distribution of the $WS_2$ triangles by edge detection and correlation of these edges with the a-plane (11-20) of the sapphire wafer edge. The rotation angles are defined clockwise with respect to the surface normal of the a-plane in the [11-20] direction, where a rotation angle of 60° is related to the [10-10] direction of the $WS_2$.[43] An example scanning electron microscope (SEM) image of this evaluation is shown in Figure 1 b). A histogram of the $WS_2$ orientations is constructed from the areas between the aforementioned wrinkles and is shown in Figure 1 c). To identify the effect of the underlying FLG on the orientation of the $WS_2$ triangles, the orientation distribution determined from the growth of $WS_2$ on the bare sapphire substrate is superimposed on the resulting histogram as a black line.[43] Both histograms clearly show the alignment of the WS$_2$ triangles to the sapphire crystal axis along [11-20] (0°) and to its analogous counterpart along [2-1-10] (60°). The difference in the abundance ratio of the 60° and 0° orientations is explained by the step-guided growth, which is more dominant on sapphire than on FLG/sapphire.[36] For the latter, the heating to a temperature of 1400 °C to grow the FLG induces an additional annealing of the sapphire, resulting in larger terraces and consequently a reduction in step-guided growth. This affects the growth of the FLG and consequently the orientation of the WS$_2$ crystallites, resulting in a more uniform distribution. Furthermore, the distribution on the FLG shows additional peaks along the [10-10] (30°) and [01-10] (-30°) directions of the sapphire. This can be related to the non-polar nature of graphene which, unlike the polar WS$_2$, results in a repetition of the hexagonal lattice every 30° compared to the observed 60° for WS$_2$, allowing the nucleation of the FLG in these orientations and subsequently the nucleation of WS$_2$ on the FLG in these orientations.

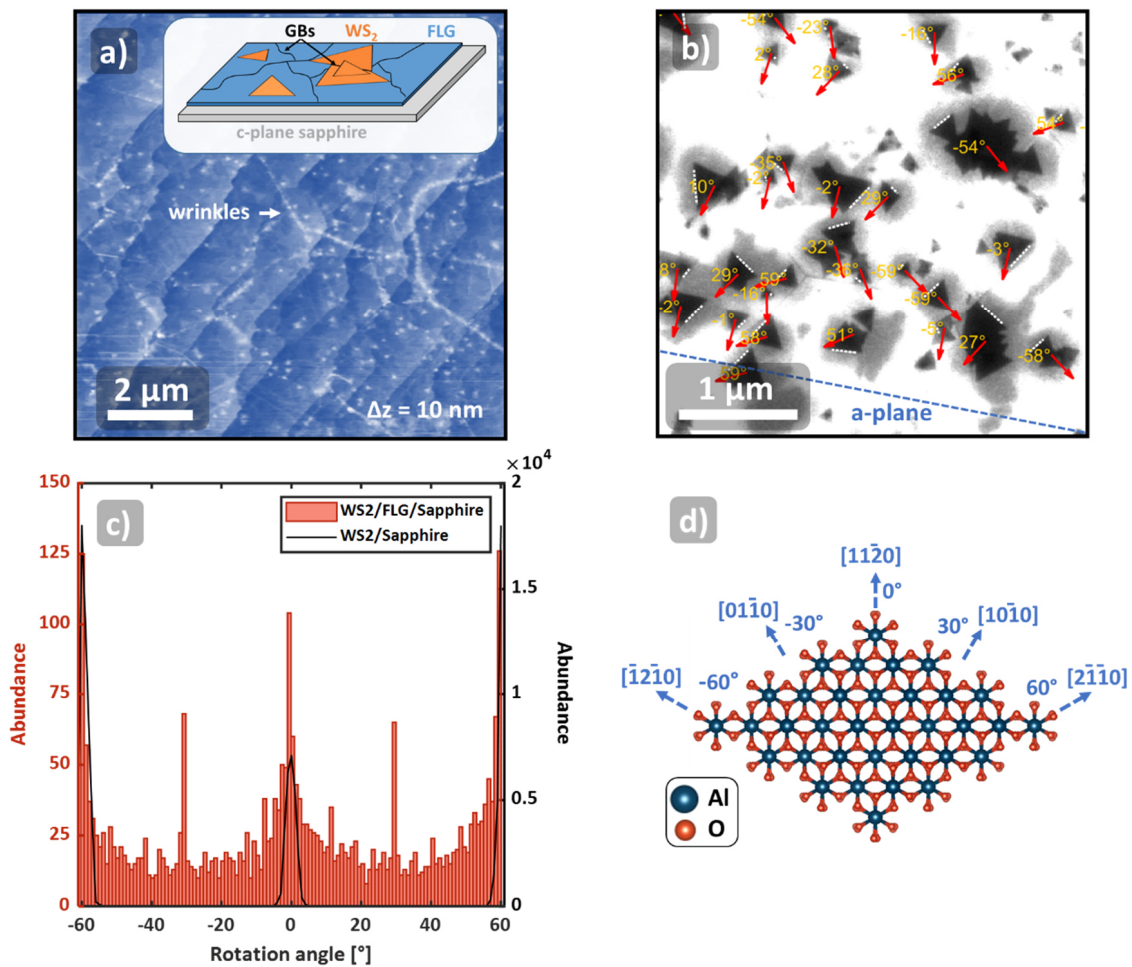

**Figure 1: Orientation analysis between the WS$_2$ triangles and the c-plane sapphire substrate.** a) AFM image of the vertical 2D heterostructure of WS$_2$/FLG on sapphire. A schematic of the sample features is added. b) Example SEM image of the vertical 2D heterostructure of WS$_2$/FLG on sapphire illustrating the edge detection scheme applied to the WS$_2$ triangles. The wafer orientation is indicated by a blue dashed line, the recognized edges by white dotted lines and the resulting rotation angles with respect to the a-plane (11-20) of the wafer are indicated by red arrows. A rotation angle of 60° refers to the [10-10] direction of the WS$_2$. c) Histogram showing the preferred orientations of the WS$_2$ triangles on the FLG on sapphire. The orientation relation of WS$_2$ grown under comparable growth conditions on the bare sapphire substrate is added for comparison (black line).[43] d) The determined crystallographic directions are illustrated in the sapphire crystal model generated by VESTA.[48]

Besides the expected broadening of the distribution around these peaks, which was found to be related to the steps on the sapphire for the growth of WS$_2$ on the bare sapphire substrate, all other orientations become available on the FLG. In addition to the 0°, 30° and 60° orientations, the FLG allows the WS$_2$ to grow in almost any other orientation, potentially allowing the growth of arbitrary rotation angles with respect to the underlying layer. By revealing the underlying mechanisms, one could promote the growth of very specific orientations and potentially engineer any desired moiré twist angle.

In the next section, we evaluate the 4D-SNBD of the transferred vertical 2D heterostructure of WS$_2$ on FLG. Here we first focus on the separation of the WS$_2$ and the FLG signal to disentangle both material contributions. With this information we reveal the relation between WS$_2$ and FLG orientations.

### 3.2. 4D Scanning Nano-Beam Diffraction Results of the Vertical Heterostructure

To disentangle the contributions of the WS$_2$ and the underlying FLG lattices we change the semi-convergence angle from 21 mrad, as used for high-resolution imaging, to 0.8 mrad for 4D-SNBD data acquisition as illustrated in Figure 2 a).

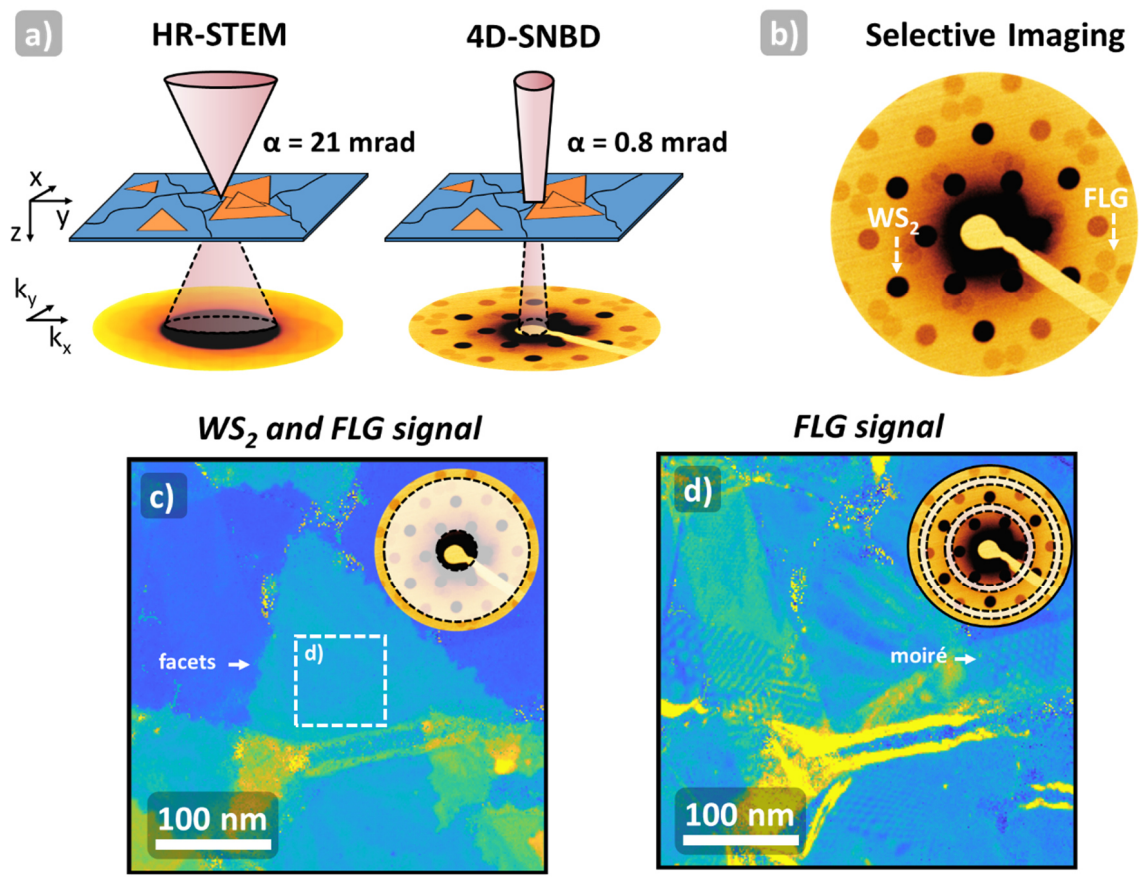

**Figure 2: High-resolution and 4D-SNBD imaging of the 2D vertical heterostructure.** a) Illustration of the HR-STEM and 4D-SNBD imaging after PMMA transfer of the WS$_2$/FLG onto a TEM grid. b) PACBED of a section of the WS$_2$/FLG heterostructure. c) The VDF image with only the direct beam removed shows a larger WS$_2$ monolayer triangle on FLG. d) The VDF image of the same region generated by selective masking based on FLG diffraction spot distances highlights the FLG structure.

Due to the difference in lattice constant between $WS_2$ and graphene the respective Bragg peaks are separated in this mode, as shown in the position-averaged (convergent) beam electron diffraction pattern (PACBED) of a representative $WS_2$ triangle in Figure 2 b). By applying an annular selection mask to the 4D data set, virtual dark field (VDF) images can be generated. A VDF omitting the direct beam shows a $WS_2$ triangle on top of the FLG (see Figure 2 c). In addition, selective masking of the FLG Bragg peaks allows the FLG to be imaged separately, revealing the nanocrystalline structure of the underlying FLG, as shown in Figure 2 d). The VDF images and the PACBED reveal an epitaxial alignment of $WS_2$ to the underlying FLG, as well as the successful PMMA transfer of the entire 2D heterostructure (cf. S2).

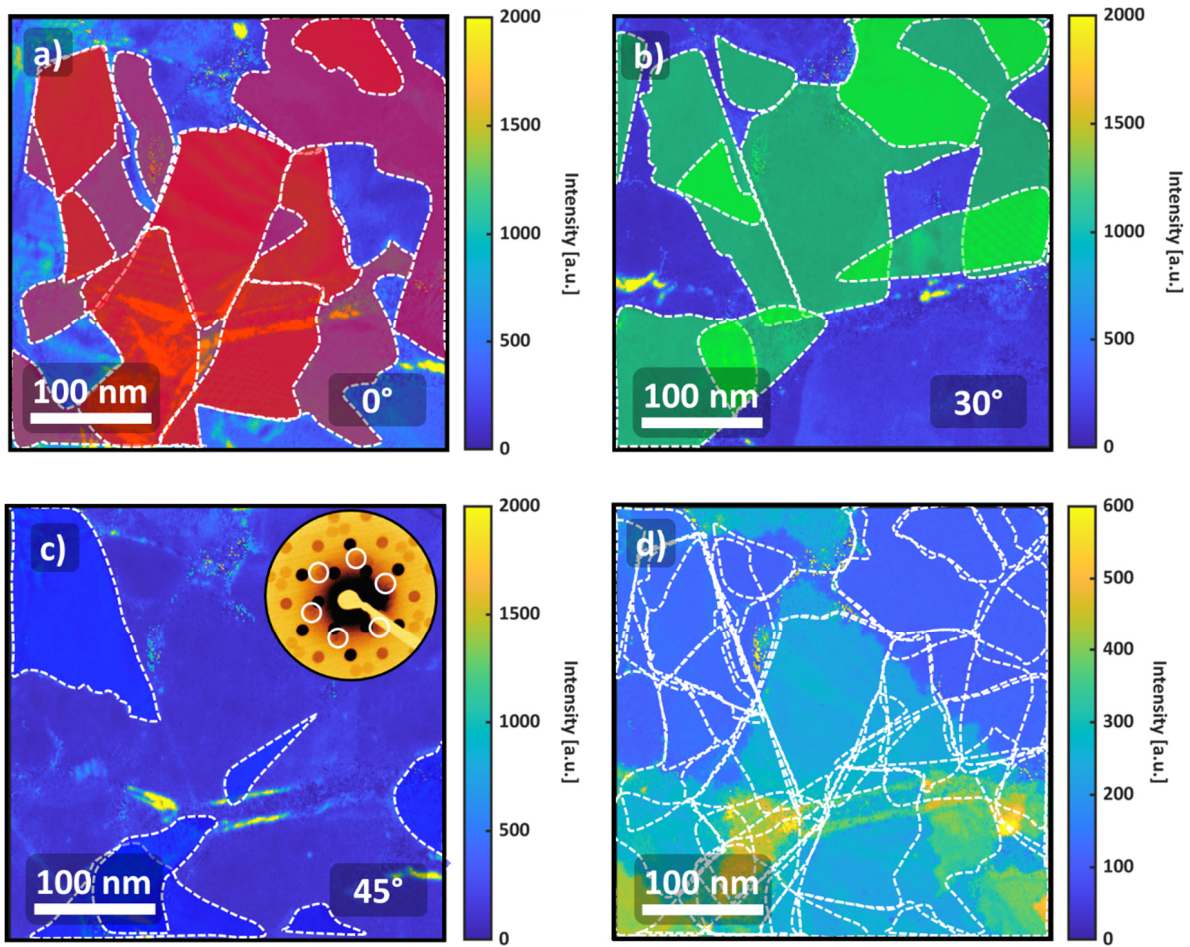

**Figure 3: Orientation analysis of the FLG and visualization of the GBs in the FLG.** a-c) VDF images of the same region as shown in Figures 2 c) and d), generated by masking the graphene Bragg peaks with a lattice rotation of 0°, 30° and 45°, respectively. An example mask is shown in the inset of c). The GBs are superimposed as white dashed lines. d) VDF image from Figure 2 c) superimposed with the extracted GBs for lattice rotations of 0°, 18°, 23°, 30°, 39° and 45°.

We use this selective masking for data sets recorded at different positions in the 2D heterostructure to investigate the orientation relationship between FLG and $WS_2$ in more detail. First, we focus on the orientation of the graphene layers. For this purpose, we generate VDF images by masking only the FLG Bragg peaks with fixed lattice rotations, as shown for example in Figure 3 a-c) and S3 for the lattice rotations of 0°, 30° and 45°. The rotation angles are determined with respect to the most common orientation of the $WS_2$ lattice. This allows us to correlate the FLG orientations with the histogram shown in Figure 1 c). From the dark field images, the number of layers in the FLG with that particular orientation is deduced. To improve the statistics, the orientation mapping is done for several areas on the sample, which is shown in S4. The orientation maps show that the graphene preferentially aligns with a rotation of 0° or 60°. In addition, a rotation of 30° is the second most abundant orientation. This implies that - to a large extent - the first graphene layer is aligning with 0°, 30° or 60° rotation to the sapphire substrate. Consequently, the $WS_2$ triangles nucleate preferentially along these orientations, resulting in the histogram observed in Figure 1 c). This correlation can also be seen by calculating the misorientation between the graphene and the $WS_2$ lattice, which is shown in S4 and which supports these findings. This shows a strong epitaxial relation between the vdW layers in the whole heterostructure. In addition, the FLG exhibits further lattice rotations, which likely form on top of the first two graphene layers. As the number of layers increases, nucleation of graphene and subsequently $WS_2$ in all other orientations becomes possible, as shown in Figure 1 c). Due to the difference in lattice constant between the 0° and 30° rotated graphene and the sapphire substrate, the nucleation of these additional orientations could be locally energetically favoured.

In the next step, we mark all GBs in the FLG with white lines in the VDF images in Figure 3 a-c). By doing this for all observed orientations of the graphene, a map of the GBs in the FLG is created. To illustrate the dependence of these GBs in the FLG on the growth of the $WS_2$, this map is superimposed on the sample region shown in Figure 2 c), as seen in Figure 3 d). Together with the orientation distributions determined, this shows the strong correlation between the $WS_2$ and the underlying FLG. We find that the boundaries of the $WS_2$ monolayer triangles coincide with the GBs in the FLG, which explains the limitation in the size of the $WS_2$ triangles. This is supported by the fact that, on the one hand, increasing the growth time from 1800 s to 3600 s does not lead to a significant increase in the size of the triangles (cf. S5), while, on the other hand, the $WS_2$ layers already coalesce at a growth time of 3600 s, when grown under the same growth conditions on the bare sapphire substrate.[42] This implies that there are regions on the FLG, as seen in the upper left and upper right regions of Figure 3 d), which do not favour the nucleation of the $WS_2$. These are regions with at least 4-5 layers of the FLG with different lattice orientations as seen in S6. We suggest that here the $WS_2$ may receive multiple epitaxy constraints from the underlying FLG resulting in no nucleation at all.

In addition, micro-facetted edges of the $WS_2$ triangles are observed as marked in Figure 2 c). These facets are formed at the GBs of the FLG, highlighting the epitaxial relation between the $WS_2$ and the FLG.

When aiming to grow a coalesced 2D heterostructure, a modification and optimization of the graphene growth parameters would be required to reduce this strong influence of the graphene GBs. However, we would like to emphasize that especially the GBs in the 2D layers play a crucial role in the formation of the moiré patterns, which we will discuss in the next section.

### 3.3 Effect of Grain Boundaries on the Formation of Moiré Patterns

In addition to the discussed influence of the GBs of the graphene, which limit the size and the growth of the $WS_2$ in certain regions of the sample, we find that the GBs play a crucial role in the formation and modification of the moiré structures in the $WS_2$ layers. Firstly, we observe the formation of a wide variety of moiré twist angles ranging from large to small twist angles in the $WS_2$ layers. At certain locations of the sample, as shown in Figure 4 a), the moiré angle can be correlated with the orientation of the underlying $WS_2$ layer, since the monolayer areas are still visible there. To analyze the rotation angles, we disentangle the contributions of the individual $WS_2$ lattices by Fourier filtering. We also superimpose the $WS_2$ crystal structure on the high-resolution image to identify the S positions (cf. S7). Since the $WS_2$ is a polar material, the angles between the grains can vary between 0 and 60°. Therefore, the rotation angles cannot only be determined from the FFTs alone and require an additional check of the lattice polarity as considered for the growth of WS$_2$ on the bare sapphire substrate.[43] Using this approach, we evaluate several regions that show moiré structures and at the same time have identifiable GBs in the lower WS$_2$ layer. One observes a linear relation between the observed moiré angle and the angle of the GB, as shown in Figure 4 b). The data points show the formation of a moiré at a GB due to the growth of one orientation over the other. These GBs appear to be essential for the formation of moiré structures in the studied MOCVD grown heterostructure. In addition, in some cases we observe that the moiré pattern also stops at the GB of the underlying WS$_2$ layer (cf. S8), indicating that the moiré patterns do not modulate arbitrarily across any GB, but only allow certain configurations. Here it is worth noting that all the observed moiré angles are close to the twist angles predicted for hexagonal lattices with a low moiré order.[49]

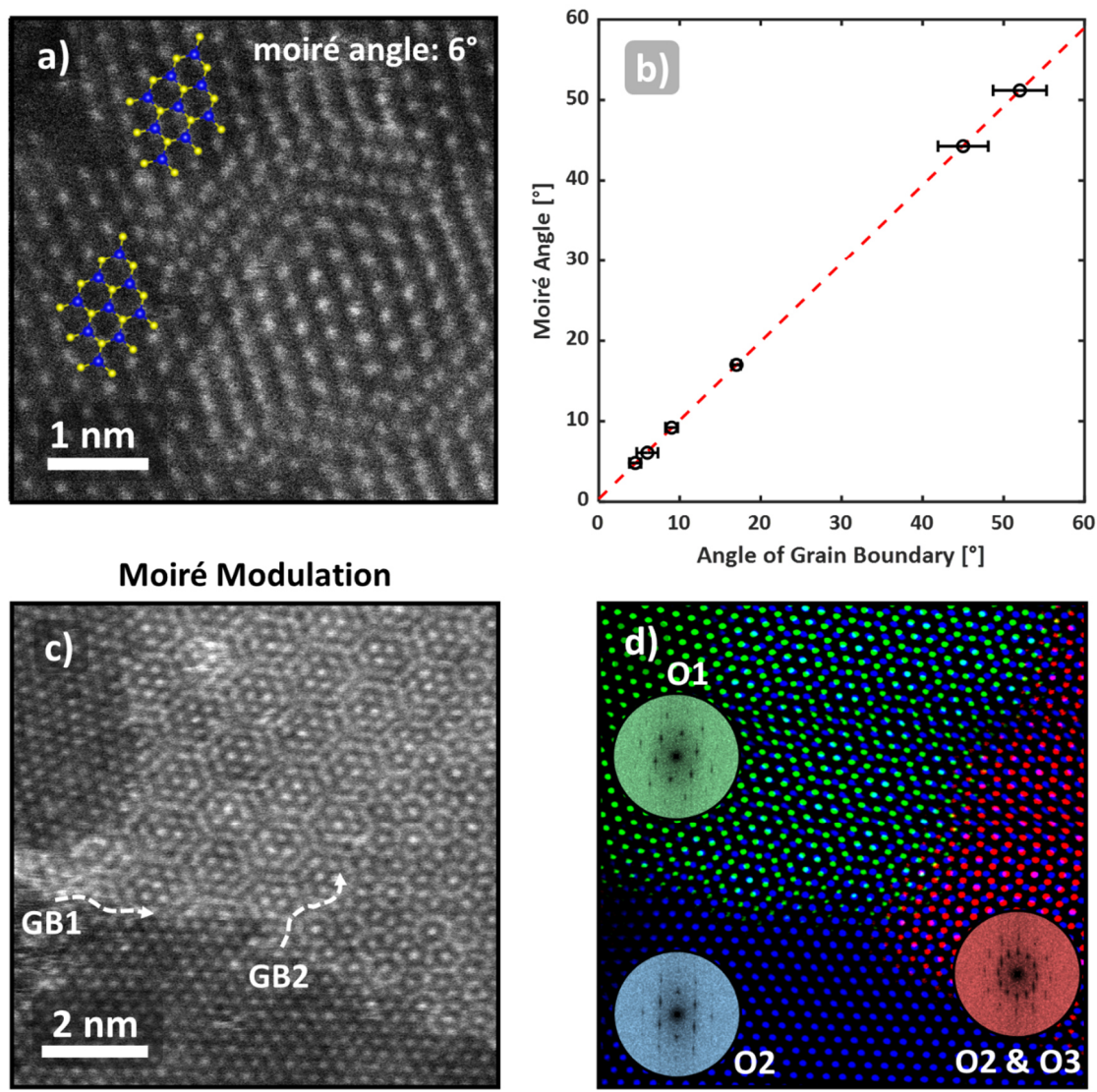

**Figure 4: Moiré modulation across GBs.** a) High-resolution ADF image of a moiré pattern formed across a 6° GB in the underlying WS$_2$ layer. b) Determined moiré angle as a function of the underlying WS$_2$ GB. c) High-resolution ADF image of a sample region showing the modulation of an already formed -30° moiré pattern to a 14° moiré pattern. The resulting moiré angle is modulated by the growth of the upper WS$_2$ layer with orientation O3 over a 44° GB between the orientations O3 and O1 in the underlying WS$_2$ layer. d) RGB image generated by Fourier filtering of the three different WS$_2$ lattice orientations. The FFTs are shown as insets.

In addition, we observe regions that show the modulation of a pre-existing moiré pattern by growth of the upper layer across a GB in the lower $WS_2$ layer, as shown in Figure 4 c). To visualize the contributions of each orientation, the $WS_2$ lattices can be separated by FFT filtering and plotted in an RBG image in Figure 4 d).

Since a spontaneous formation of the 2$^{nd}$ $WS_2$ layer on top of the first layer with a different orientation is unlikely, as also observed at different locations on the sample (cf. S9), we conclude that the boundary is located in the lower layer and that the upper $WS_2$ layer with orientation 2 (blue) grows over this boundary. Consequently, we identify GB1 between the 1$^{st}$ (green) and the 2$^{nd}$ (blue) orientations. GB2 is formed between the 1$^{st}$ (green) and the 3$^{rd}$ (red) orientation. Thus, we identify here a GB of 44° between the 1$^{st}$ (green) and the 3$^{rd}$ (red) orientation. This leads to a modification of the pre-existing moiré angle of -30° to a moiré angle of 14°, showing the modulation of this moiré pattern across a GB in the underlying $WS_2$ layer. Furthermore, we find that growth over polar GBs, which do not exhibit twist but a change in the lattice polarity, leads to a change in stacking order in bilayer $WS_2$ (cf. S10).

We believe that the observed modulation of the moiré structure due to the presence of underlying GBs may provide a controllable pathway for the growth of moiré patterns with a defined twist angle. Precise tuning of specific GBs, e.g., as a function of the grain mismatch, is desirable.

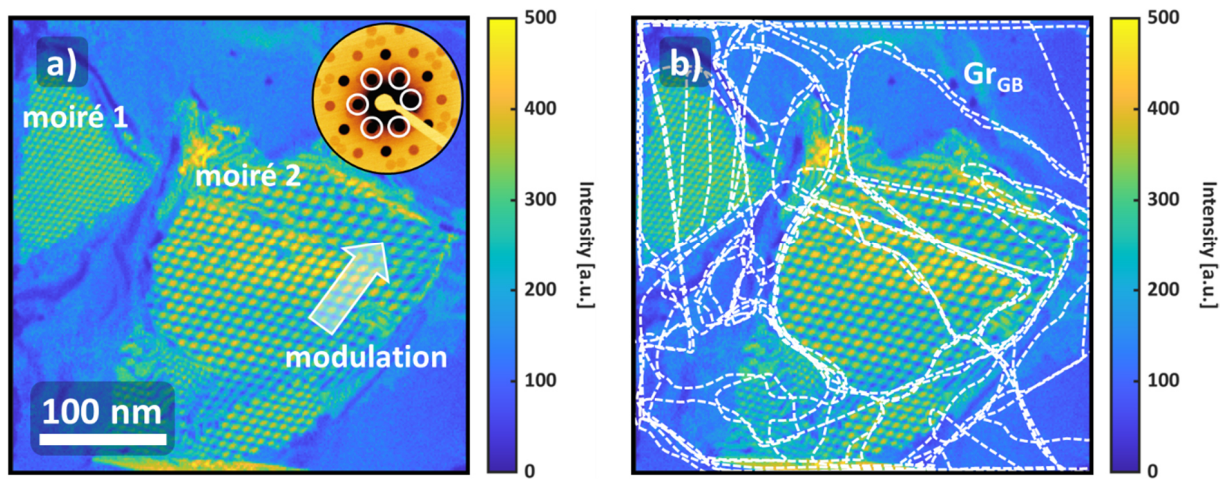

**Figure 5: Moiré modulation across GBs of the underlying FLG.** a) Virtual dark field image of a region containing two low-angle $WS_2$ moiré patterns generated by masking only the first order of diffraction of $WS_2$ for a fixed rotation angle of 0°, as shown in the inset. Due to the low twist angle of 2.7° and 1.7° in the moiré regions 1 and 2, respectively, an intensity fluctuation caused by the moiré lattice can be seen. b) The additional modulation of the moiré lattice coincides with the GBs in the underlying FLG layers, which are superimposed by white lines.

Finally, in the 4D-SNBD data sets, we observe an influence of the underlying FLG layers on the moiré pattern formed between the $WS_2$ layers. The periodic lattice repetition of the low twist angle moiré pattern is shown as intensity modulation in the VDF images due to the change between AA and AB like stacking regions. Within the given resolution of about 1-2 nm, the low angle moiré patterns can be imaged, as already seen for the multilayer regions of the FLG in Figure 2 d). Similarly, by masking the Bragg peaks close to the orientations of the two $WS_2$ layers, a virtual darkfield image sensitive to the moiré lattice can be obtained. An example region showing two moiré structures with a twist angle of 2.7° and 1.7° respectively is shown in Figure 5 a) and the high-resolution image of moiré 2 is shown in S11. Both moiré patterns form with slight variations around the 0° orientation, which we correlate with the [11-20] direction of the sapphire. The first moiré pattern shows a rather constant pattern ranging over an area of about 100 nm². In contrast, the second area shows a dominant moiré pattern but also additional local variations within the flake. Using the same approach as illustrated in Figure 3, we determine the GBs of the underlying FLG and superimpose them on the image, as illustrated in Figure 5 b). Here, we observe a strong correlation between the GBs in the FLG and the $WS_2$ moiré pattern, which convincingly explains the modulation of the pattern within the flake. We believe that the modulation originates from a modulation in the lower $WS_2$ layer, as this should be more strongly influenced by the underlying FLG layers. Similarly, the upper $WS_2$ layer is less modulated due to the weak vdW bond, leading to the overall modulation in the moiré pattern.

This finding is further supported by an analysis of the FLG below moiré region 1, where two low-angle moiré lattices, had formed prior to $WS_2$ growth (cf. S12). We observe that this small distortion leads to an almost unaffected alignment between the FLG and the $WS_2$, resulting in a homogeneous moiré pattern. The opposite case is found in region 2. Here, one of the FLG layers below the main moiré region exhibits a strong change in lattice rotation, leading to a large angle GB in the FLG, which consequently strongly affects the orientation of the $WS_2$.

Finally, we would like to emphasize that in bottom-up synthesis, the formation of moiré patterns clearly depends on the orientation of the underlying layer in the vertical 2D heterostructure when grown on a substrate that allows for an epitaxial relation to the 2D layer. In particular, the GBs in the lower layer lead to the formation of moiré patterns by the growth of the upper layer across these boundaries. In addition, the moiré patterns can be further modulated by growth across adjacent GBs. In particular, the addition of FLG in these heterostructures allows the growth of low-angle moiré patterns, which are of fundamental and technological interest. Overall, this type of bottom-up synthesis is urgently needed for large-scale production, as opposed to simple exfoliation and mechanical stacking.

## 4. Discussion and Conclusion

In this paper we study a complex 2D heterostructure of $WS_2$ and FLG grown by MOCVD on c-plane sapphire. In a first step we determine the orientation of the $WS_2$ with respect to the sapphire substrate by edge detection of the triangles observed in the SEM images. To identify the orientation relation of the $WS_2$ to the underlying FLG, the heterostructure is analyzed using the capabilities of 4D STEM. 4D-SNBD is used to separate the signals from $WS_2$ and FLG in the 2D heterostructures. The VDF images, generated with specific selective masking of the Bragg diffraction spots, reveal the structure of both $WS_2$ and FLG at the nanometer scale and allow for separate orientation mapping of these materials. While the growth of $WS_2$ on the bare sapphire substrate only allows an orientation with respect to the [11-20] or [2-1-10] and analogous directions, the growth on FLG/sapphire shows a broader distribution, covering all possible rotations. Cumulation still occurs along the [11-20] or [2-1-10] directions, as well as the [01-10] and [10-10] directions, but to a lesser extent. Despite the complexity of the specific structure, we can identify a mechanism for modifying the $WS_2$ triangle orientations that is not possible on pristine sapphire substrates.

We show that 4D-SNBD is a useful technique to visualize the formation of GBs in the underlying graphene layers. Our results explain the size extension and size limitation of $WS_2$ triangles. The growth of larger $WS_2$ flakes or even a coalesced film may be enabled by a reduction of GBs in FLG. Furthermore, our analysis of the moiré structures formed between $WS_2$ layers shows a strong dependence on the GBs of the lower layers. We find that these

GBs lead both to modulation of existing moiré features and to the formation of new patterns. The situation in the whole heterostructure is shown to be even more complex, since the FLG can influence all subsequent moiré patterns if a large-angle GB is present.

Our results suggest that if one can tune the growth of GBs in the underlying $WS_2$ and/or graphene layer, one obtains a controllable parameter for growth of moiré structures with specific twist angles. In particular, substrate modification and pre-treatment have been reported, e.g., for the growth of single-crystal graphene on Cu(111) or Ge(110) substrates on a wafer scale.[7,17,19,29] Similarly, the realization of a patterned substrate or the specific preparation of single or double-atomic steps on the surface[50,51] could allow the control of the nucleation of defined graphene domains, leading to specific GBs. Other approaches, such as post-growth manipulation of the graphene, have already been used in the literature to realize unidirectional 2D materials[52] that could be modified to control the growth of GBs.

We believe that our findings provide an additional approach to twist angle control for the synthesis of 2D materials via MOCVD and comparable growth methods. In particular, low twist angle moiré patterns are observed that are not found in samples grown directly on sapphire and which are of particular interest due to their predicted novel physical properties.[2,7,12]

**Acknowledgements**

This work was supported by the German Research Foundation in the framework of the Collaborative Research Centre SFB 1083 "Structure and Dynamics of Internal Interfaces" (project number: 223848855).

**Conflict of Interest**

The authors have no financial or commercial Conflict of Interest.

**Data Availability Statement**

The 4D data that support the findings of this study are openly available in [repository name e.g "figshare"] at http://doi.org/[doi], reference number [reference number].

**References:**

[1] K. S. Novoselov, A. K. Geim, S. V. Morozov, D. Jiang, Y. Zhang, S. V. Dubonos, I. V. Grigorieva, A. A. Firsov, *Science (New York, N.Y.)* **2004**, *306*, 666.

[2] K. S. Novoselov, A. Mishchenko, A. Carvalho, A. H. Castro Neto, *Science (New York, N.Y.)* **2016**, *353*, aac9439.

[3] K. S. Novoselov, D. Jiang, F. Schedin, T. J. Booth, V. V. Khotkevich, S. V. Morozov, A. K. Geim, *Proceedings of the National Academy of Sciences of the United States of America* **2005**, *102*, 10451.

[4] M. Jonson, *THE ROYAL SWEDISH ACADEMY OF SCIENCES - Nobel Prize in Physics 2010* **2010**.

[5] M. Z. Hasan, C. L. Kane, *Rev. Mod. Phys.* **2010**, *82*, 3045.

[6] S. Tang, A. Grundmann, H. Fiadziushkin, A. Ghiami, M. Heuken, A. Vescan, H. Kalisch, *MRS Advances* **2022**, *7*, 751.

[7] S. Zhang, J. Liu, M. M. Kirchner, H. Wang, Y. Ren, W. Lei, *J. Phys. D: Appl. Phys.* **2021**, *54*, 433001.

[8] C. R. Dean, L. Wang, P. Maher, C. Forsythe, F. Ghahari, Y. Gao, J. Katoch, M. Ishigami, P. Moon, M. Koshino, T. Taniguchi, K. Watanabe, K. L. Shepard, J. Hone, P. Kim, *Nature* **2013**, *497*, 598.

[9] M. Yankowitz, J. Xue, D. Cormode, J. D. Sanchez-Yamagishi, K. Watanabe, T. Taniguchi, P. Jarillo-Herrero, P. Jacquod, B. J. LeRoy, *Nature Phys* **2012**, *8*, 382.

[10] B. Hunt, J. D. Sanchez-Yamagishi, A. F. Young, M. Yankowitz, B. J. LeRoy, K. Watanabe, T. Taniguchi, P. Moon, M. Koshino, P. Jarillo-Herrero, R. C. Ashoori, *Science (New York, N.Y.)* **2013**, *340*, 1427.

[11] K. L. Seyler, P. Rivera, H. Yu, N. P. Wilson, E. L. Ray, D. G. Mandrus, J. Yan, W. Yao, X. Xu, *Nature* **2019**, *567*, 66.

[12] K. Tran, G. Moody, F. Wu, X. Lu, J. Choi, K. Kim, A. Rai, D. A. Sanchez, J. Quan, A. Singh, J. Embley, A. Zepeda, M. Campbell, T. Autry, T. Taniguchi, K. Watanabe, N. Lu, S. K. Banerjee, K. L. Silverman, S. Kim, E. Tutuc, L. Yang, A. H. MacDonald, X. Li, *Nature* **2019**, *567*, 71.

[13] Y. Pan, S. Fölsch, Y. Nie, D. Waters, Y.-C. Lin, B. Jariwala, K. Zhang, K. Cho, J. A. Robinson, R. M. Feenstra, *Nano letters* **2018**, *18*, 1849.

[14] E. M. Alexeev, D. A. Ruiz-Tijerina, M. Danovich, M. J. Hamer, D. J. Terry, P. K. Nayak, S. Ahn, S. Pak, J. Lee, J. I. Sohn, M. R. Molas, M. Koperski, K. Watanabe, T. Taniguchi, K. S. Novoselov, R. V. Gorbachev, H. S. Shin, V. I. Fal'ko, A. I. Tartakovskii, *Nature* **2019**, *567*, 81.

[15] Y. Beckmann, A. Grundmann, L. Daniel, M. Abdelbaky, C. McAleese, X. Wang, B. Conran, S. Pasko, S. Krotkus, M. Heuken, H. Kalisch, A. Vescan, W. Mertin, T. Kümmell, G. Bacher, *ACS applied materials & interfaces* **2022**, *14*, 35184.

[16] A. K. Geim, I. V. Grigorieva, *Nature* **2013**, *499*, 419.

[17] J. D. Caldwell, T. J. Anderson, J. C. Culbertson, G. G. Jernigan, K. D. Hobart, F. J. Kub, M. J. Tadjer, J. L. Tedesco, J. K. Hite, M. A. Mastro, R. L. Myers-Ward, C. R. Eddy, P. M. Campbell, D. K. Gaskill, *ACS nano* **2010**, *4*, 1108.

[18] M.-Y. Li, C.-H. Chen, Y. Shi, L.-J. Li, *Materials Today* **2016**, *19*, 322.

[19] B. V. Lotsch, *Annu. Rev. Mater. Res.* **2015**, *45*, 85.

[20] F. Pizzocchero, L. Gammelgaard, B. S. Jessen, J. M. Caridad, L. Wang, J. Hone, P. Bøggild, T. J. Booth, *Nature communications* **2016**, *7*, 11894.

[21] F. Zhang, C. Erb, L. Runkle, X. Zhang, N. Alem, *Nanotechnology* **2018**, *29*, 25602.

[22] A. Koma, K. Sunouchi, T. Miyajima, *Microelectronic Engineering* **1984**, *2*, 129.

[23] A. Koma, K. Sunouchi, T. Miyajima, in *Proceedings of the 17th International Conference on the Physics of Semiconductors* (Eds.: J. D. Chadi, W. A. Harrison), Springer New York. New York, NY **1985**, p. 1465.

[24] Z. Zuo, Z. Xu, R. Zheng, A. Khanaki, J.-G. Zheng, J. Liu, *Scientific reports* **2015**, *5*, 14760.

[25] J. M. Wofford, S. Nakhaie, T. Krause, X. Liu, M. Ramsteiner, M. Hanke, H. Riechert, J. M. J Lopes, *Scientific reports* **2017**, *7*, 43644.

[26] E. Xenogiannopoulou, P. Tsipas, K. E. Aretouli, D. Tsoutsou, S. A. Giamini, C. Bazioti, G. P. Dimitrakopulos, P. Komninou, S. Brems, C. Huyghebaert, I. P. Radu, A. Dimoulas, *Nanoscale* **2015**, *7*, 7896.

[27] K. E. Aretouli, P. Tsipas, D. Tsoutsou, J. Marquez-Velasco, E. Xenogiannopoulou, S. A. Giamini, E. Vassalou, N. Kelaidis, A. Dimoulas, *Applied Physics Letters* **2015**, *106*.

[28] S. Vishwanath, X. Liu, S. Rouvimov, L. Basile, N. Lu, A. Azcatl, K. Magno, R. M. Wallace, M. Kim, J.-C. Idrobo, J. K. Furdyna, D. Jena, H. G. Xing, *Journal of Materials Research* **2016**, *31*, 900.

[29] Y. Gong, J. Lin, X. Wang, G. Shi, S. Lei, Z. Lin, X. Zou, G. Ye, R. Vajtai, B. I. Yakobson, H. Terrones, M. Terrones, B. K. Tay, J. Lou, S. T. Pantelides, Z. Liu, W. Zhou, P. M. Ajayan, *Nature materials* **2014**, *13*, 1135.
[30] J. Yu, J. Li, W. Zhang, H. Chang, *Chemical science* **2015**, *6*, 6705.
[31] M. Li, Y. Zhu, T. Li, Y. Lin, H. Cai, S. Li, H. Ding, N. Pan, X. Wang, *Inorg. Chem. Front.* **2018**, *5*, 1828.
[32] W. Yang, G. Chen, Z. Shi, C.-C. Liu, L. Zhang, G. Xie, M. Cheng, D. Wang, R. Yang, D. Shi, K. Watanabe, T. Taniguchi, Y. Yao, Y. Zhang, G. Zhang, *Nature materials* **2013**, *12*, 792.
[33] M. Wang, S. K. Jang, W.-J. Jang, M. Kim, S.-Y. Park, S.-W. Kim, S.-J. Kahng, J.-Y. Choi, R. S. Ruoff, Y. J. Song, S. Lee, *Advanced materials (Deerfield Beach, Fla.)* **2013**, *25*, 2746.
[34] L. Fu, Y. Sun, N. Wu, R. G. Mendes, L. Chen, Z. Xu, T. Zhang, M. H. Rümmeli, B. Rellinghaus, D. Pohl, L. Zhuang, *ACS nano* **2016**, *10*, 2063.
[35] S. M. Eichfeld, V. O. Colon, Y. Nie, K. Cho, J. A. Robinson, *2D Materials* **2016**, *3*, 25015.
[36] M. Chubarov, T. H. Choudhury, D. R. Hickey, S. Bachu, T. Zhang, A. Sebastian, A. Bansal, H. Zhu, N. Trainor, S. Das, M. Terrones, N. Alem, J. M. Redwing, *ACS nano* **2021**, *15*, 2532.
[37] A. Grundmann, C. McAleese, B. R. Conran, A. Pakes, D. Andrzejewski, T. Kümmell, G. Bacher, K. B. K. Teo, M. Heuken, H. Kalisch, A. Vescan, *MRS Advances* **2020**, *5*, 1625.
[38] N. Mishra, S. Forti, F. Fabbri, L. Martini, C. McAleese, B. R. Conran, P. R. Whelan, A. Shivayogimath, B. S. Jessen, L. Buß, J. Falta, I. Aliaj, S. Roddaro, J. I. Flege, P. Bøggild, K. B. K. Teo, C. Coletti, *Small (Weinheim an der Bergstrasse, Germany)* **2019**, *15*, e1904906.
[39] Y. Cao, V. Fatemi, A. Demir, S. Fang, S. L. Tomarken, J. Y. Luo, J. D. Sanchez-Yamagishi, K. Watanabe, T. Taniguchi, E. Kaxiras, R. C. Ashoori, P. Jarillo-Herrero, *Nature* **2018**, *556*, 80.
[40] Y. Cao, V. Fatemi, S. Fang, K. Watanabe, T. Taniguchi, E. Kaxiras, P. Jarillo-Herrero, *Nature* **2018**, *556*, 43.
[41] G. Iannaccone, F. Bonaccorso, L. Colombo, G. Fiori, *Nature nanotechnology* **2018**, *13*, 183.
[42] H. Tang, S. Pasko, S. Krotkus, T. Anders, C. Wockel, J. Mischke, X. Wang, B. Conran, C. McAleese, K. Teo, S. Banerjee, H. M. Silva, P. Morin, I. Asselberghs, A. Ghiami, A. Grundmann, S. Tang, H. Fiadziushkin, H. Kalisch, A. Vescan, S. El Kazzi, A. Marty, D. Dosenovic, H. Okuno, L. Le Van-Jodin, M. Heuken, *Journal of Crystal Growth* **2023**, *608*, 127111.
[43] D. Dosenovic, S. Dechamps, C. Vergnaud, S. Pasko, S. Krotkus, M. Heuken, L. Genovese, J.-L. Rouviere, M. den Hertog, L. Le Van-Jodin, M. Jamet, A. Marty, H. Okuno, *2D Materials* **2023**, *10*, 45024.
[44] L. Duschek, P. Kükelhan, A. Beyer, S. Firoozabadi, J. O. Oelerich, C. Fuchs, W. Stolz, A. Ballabio, G. Isella, K. Volz, *Ultramicroscopy* **2019**, *200*, 84.
[45] A. Beyer, M. S. Munde, S. Firoozabadi, D. Heimes, T. Grieb, A. Rosenauer, K. Müller-Caspary, K. Volz, *Nano letters* **2021**, *21*, 2018.
[46] N. P. Kazmierczak, M. van Winkle, C. Ophus, K. C. Bustillo, S. Carr, H. G. Brown, J. Ciston, T. Taniguchi, K. Watanabe, D. K. Bediako, *Nature materials* **2021**, *20*, 956.
[47] H. Ryll, M. Simson, R. Hartmann, P. Holl, M. Huth, S. Ihle, Y. Kondo, P. Kotula, A. Liebel, K. Müller-Caspary, A. Rosenauer, R. Sagawa, J. Schmidt, H. Soltau, L. Strüder, *Journal of Instrumentation* **2016**, *11*, P04006.

[48] K. Momma, F. Izumi, *J Appl Crystallogr* **2011**, *44*, 1272.
[49] M. Feuerbacher, *arXiv* **2020**, *2007.03542*.
[50] A. Beyer, J. Ohlmann, S. Liebich, H. Heim, G. Witte, W. Stolz, K. Volz, *Journal of Applied Physics* **2012**, *111*.
[51] K. Volz, A. Beyer, W. Witte, J. Ohlmann, I. Németh, B. Kunert, W. Stolz, *Journal of Crystal Growth* **2011**, *315*, 37.
[52] W. Yao, B. Wu, Y. Liu, *ACS nano* **2020**, *14*, 9320.

**Supporting Information**

S1: Attached presentation about the data evaluation of the 4D datasets - "PostProcessingScheme.pdf"

Description of the post processing scheme applied to the 4D-SNBD data:

First, standard correction schemes of the individual frames, including a gain, a dark frame offset, a noise threshold and out-of-time event correction are applied. The 4D diffraction data is then further optimized in the following steps. First, the center of the averaged diffraction pattern is determined by the intersection of two linear fits to four points from the 1$^{st}$ and 2$^{nd}$ diffraction order of the reciprocal $WS_2$ lattice. The positions of the four points are optimized by the center of mass (COM) of their respective diffraction discs. Knowing the center, the remaining points of the reciprocal hexagonal $WS_2$ lattice can be constructed by their expected relative positions with a rotation and scaling factor as input. The resulting positions of the reciprocal hexagonal lattice points are further optimized by the built in MATLAB contour plot and a circle fit.[1] The final positions of the reciprocal hexagonal lattice are then a simple affine transformation of the original reciprocal hexagonal lattice points. The corresponding matrix of this affine transformation is then stored and used to construct the reciprocal lattice points of the $WS_2$ and the graphene lattice. The latter includes just an additional scaling factor. With these both lattices the diffraction patterns can be separated by simple binary masks around each reciprocal lattice point. These masks are then rotated between 0° to 60°, which, for the given hexagonal symmetry, accounts for the possible rotational alignment between both lattices. As further improvement, the affine transformations can be optimized for each individual frame, leading to a determination of the shift of the center of each diffraction pattern as done in ref.[2] This shift is expected due to unintentional beam tilt while scanning and the lack of a proper correction of this de-scan within the microscope used. To reduce the computational resources needed, the beam shift is determined from a binned dataset (8x8 px) and then interpolated to the full (256x256 px) dataset. This gives a convenient correction of the de-scan. On top of this, each diffraction pattern exhibits undesirable background scattering. To account for this, a Lorentzian function is fitted to the normalized radial intensity for each diffraction pattern. To neglect the contribution of the individual Bragg reflections, only the local minima of the curve are used for the fitting. From this fit, the 2D background-signal can be interpolated and subtracted from each diffraction pattern. The resulting corrected 4D dataset is then stored in the netCDF data format.

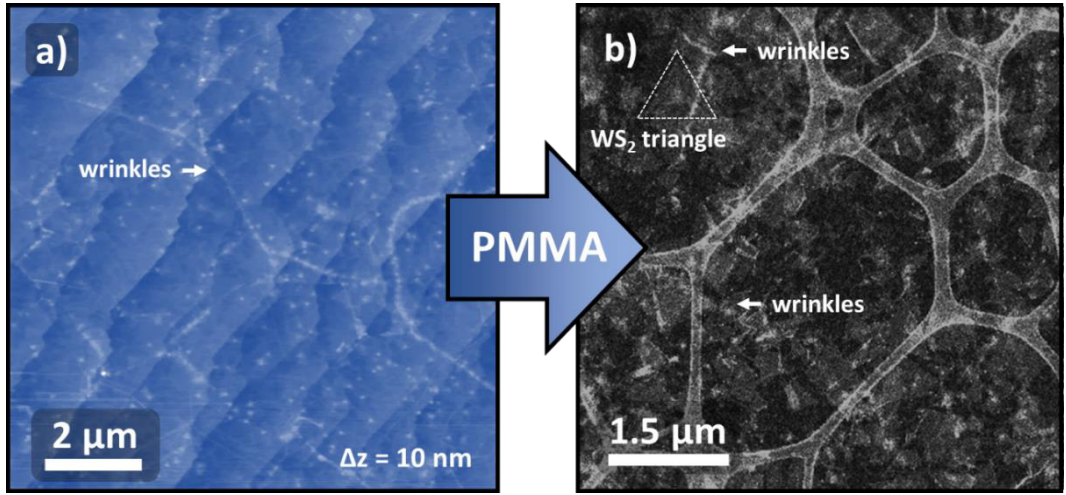

S2: a) AFM image of the vertical 2D heterostructure of $WS_2$/FLG on sapphire. b) ADF STEM image of the heterostructure after the PMMA transfer to a lacey carbon TEM grid. All sample features are preserved in the PMMA transferred samples and are marked in the image.

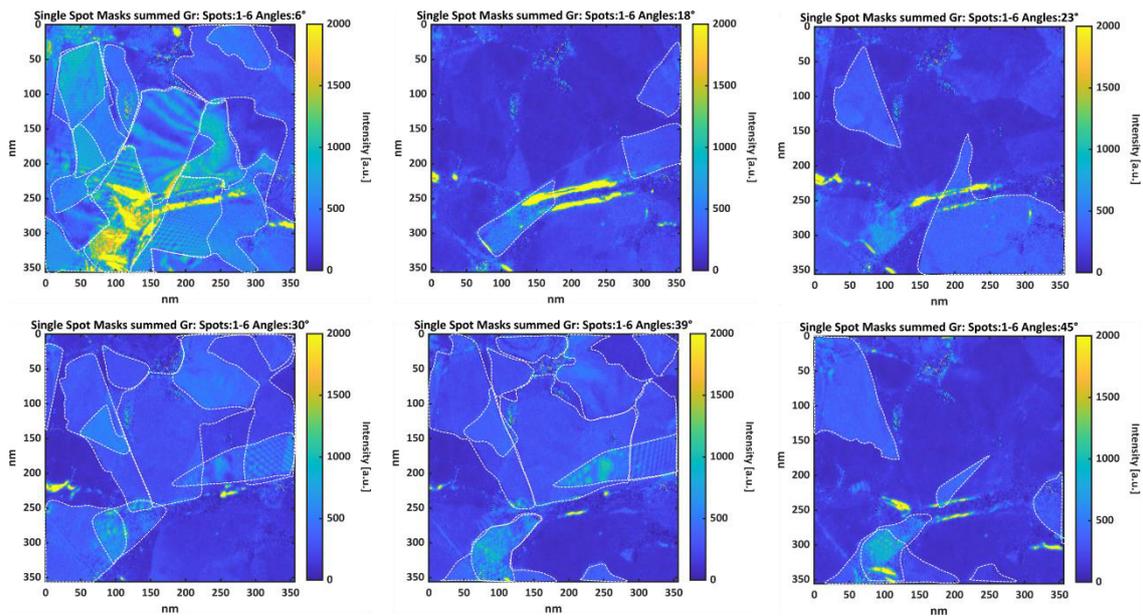

S3: Virtual dark field images created from the 4D datasets by masking the first diffraction order of the Bragg reflections of the graphene for the different rotation angles of 0°, 18°, 23°, 30° and 45°. The used mask is illustrated in Figure 3 of the main text. The rotation angles are determined with respect to the most abundant orientation of the $WS_2$. Since the rotation of the $WS_2$ to the sapphire substrate is determined in Figure 2, the rotation angle can be related to the [-2110] direction of the sapphire. In the virtual dark field images, the flake and grain boundaries in the FLG layers are marked by white lines.

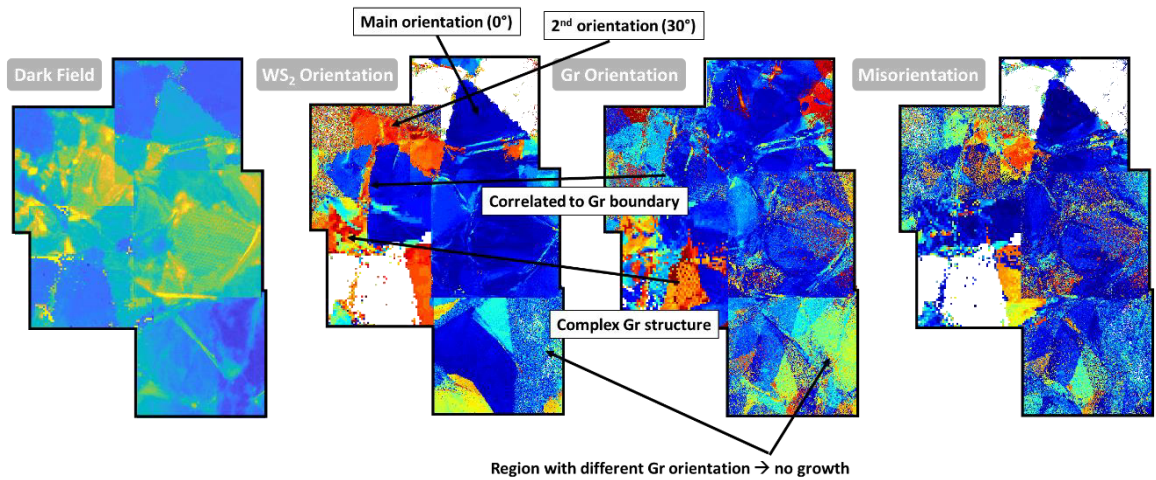

S4: Background corrected virtual dark field images generated from multiple 4D datasets obtained next to the sample region shown in Figure 3d in the main text (left). Generated orientation maps of the first $WS_2$ and Gr layers, respectively (center). The orientation maps show the two main orientations of 0° and 30°, correlating well to the observed distribution from the edge detection of the SEM images. In the STEM data, the orientation of the graphene becomes accessible, which correlates nicely to the $WS_2$ and therefore to the c-plane sapphire. This becomes even more apparent in the misorientation map (right). Smaller patches of graphene with altered orientation are seen.

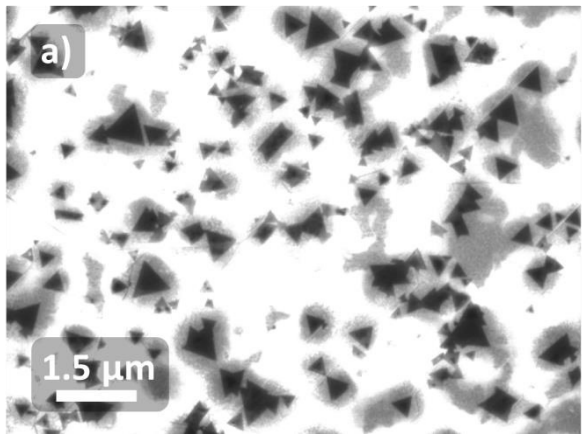
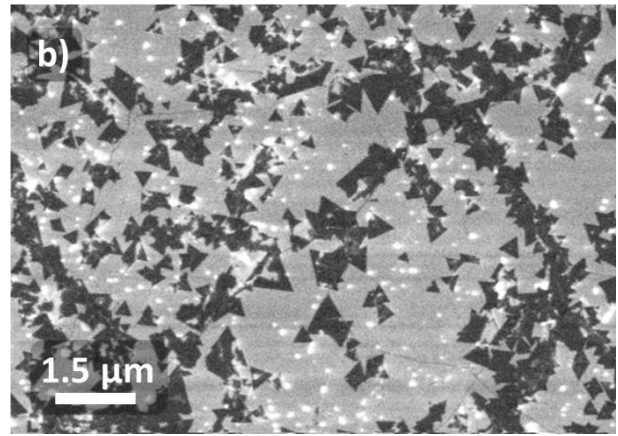

S5: SEM images of the investigated 2D heterostructures of $WS_2$ on FLG. By an increase of the growth time of the $WS_2$ from 1800s (left) to 3300s (right) no significant increase of the size of the $WS_2$ triangles is observed.

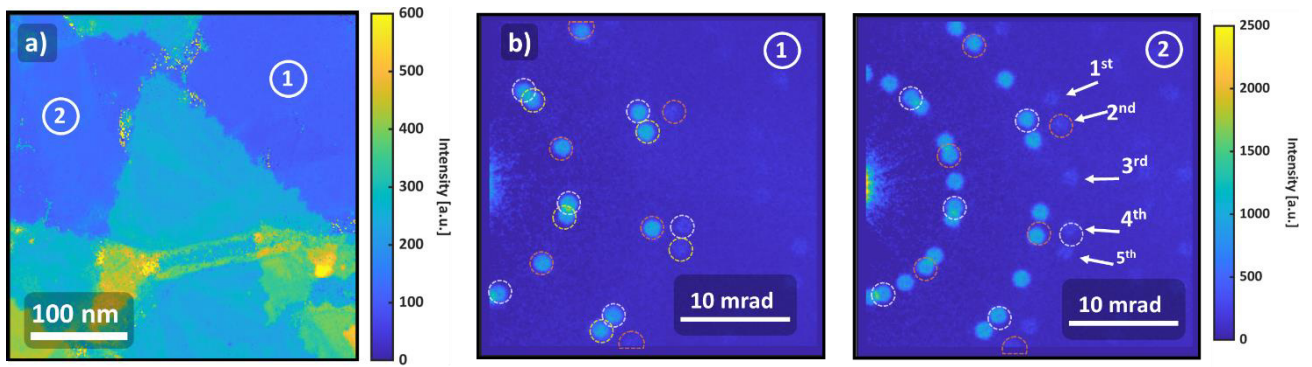

S6: a) Virtual dark field image from Figure 3 d) of the main text. b-c) Selected averaged diffraction patterns of 16x16 px from regions 1 and 2, respectively. The main rotation angles of 0° and 30° are marked by orange and white circles, respectively. Additional orientations are labeled or marked in different color. Both regions exhibit multiple graphene orientations, potentially hindering the nucleation of the $WS_2$ in these regions.

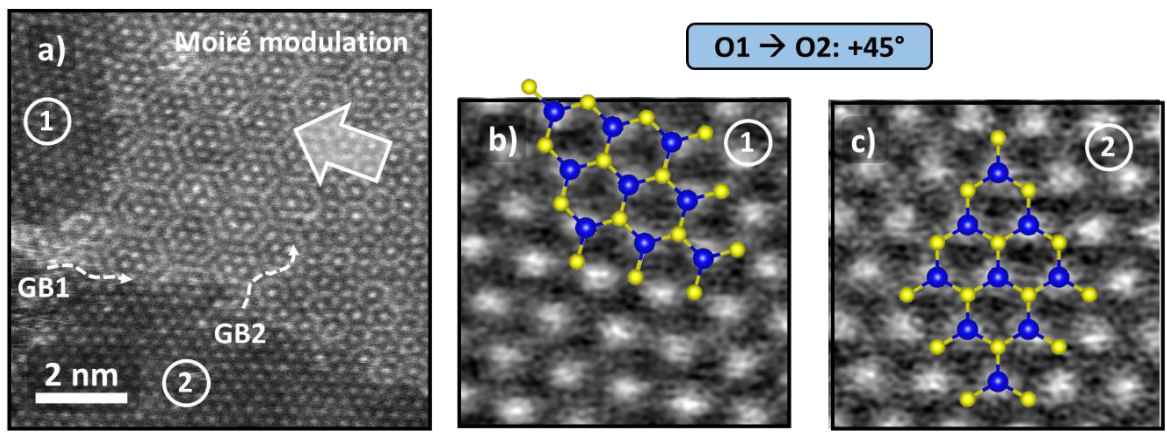

S7: a) High-resolution ADF image from Figure 4 a) of the main text showing the modulation of the moiré pattern across a grain boundary of the underlaying $WS_2$ layer. b-c) Structural model of the $WS_2$ monolayer overlaid on to the regions 1 and 2 to fit both the Tungsten (blue) and Sulphur (yellow) positions. From the structural models a lattice rotation of 45° between both monolayer areas is determined.

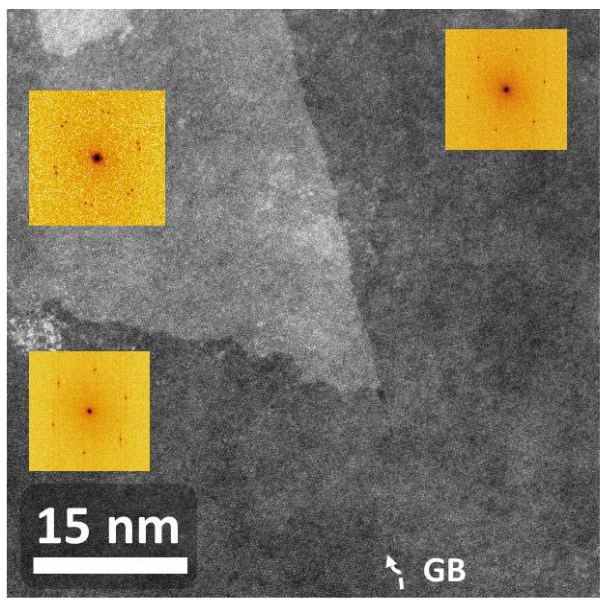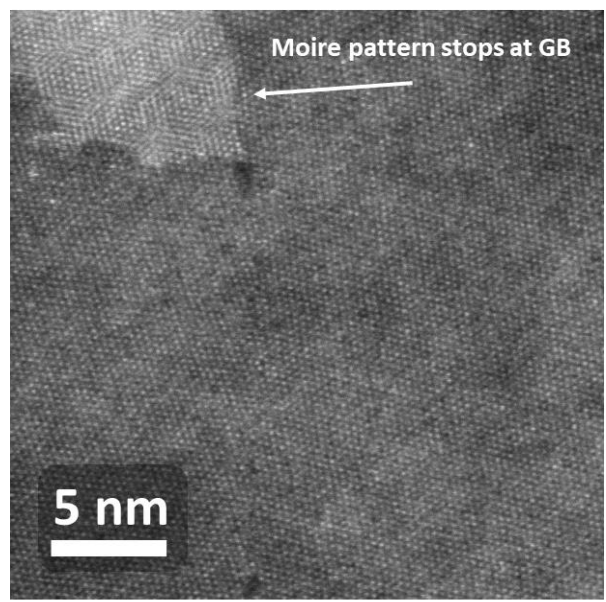

S8: High-resolution HAADF images of a $WS_2$ moiré pattern. The growth of the moiré pattern stops right at the grain boundary in the underlying $WS_2$ layer. The FFTs showing the two monolayer orientations as well as the moiré area are added as insets (left). Magnified region of the grain boundary and the moiré pattern (right).

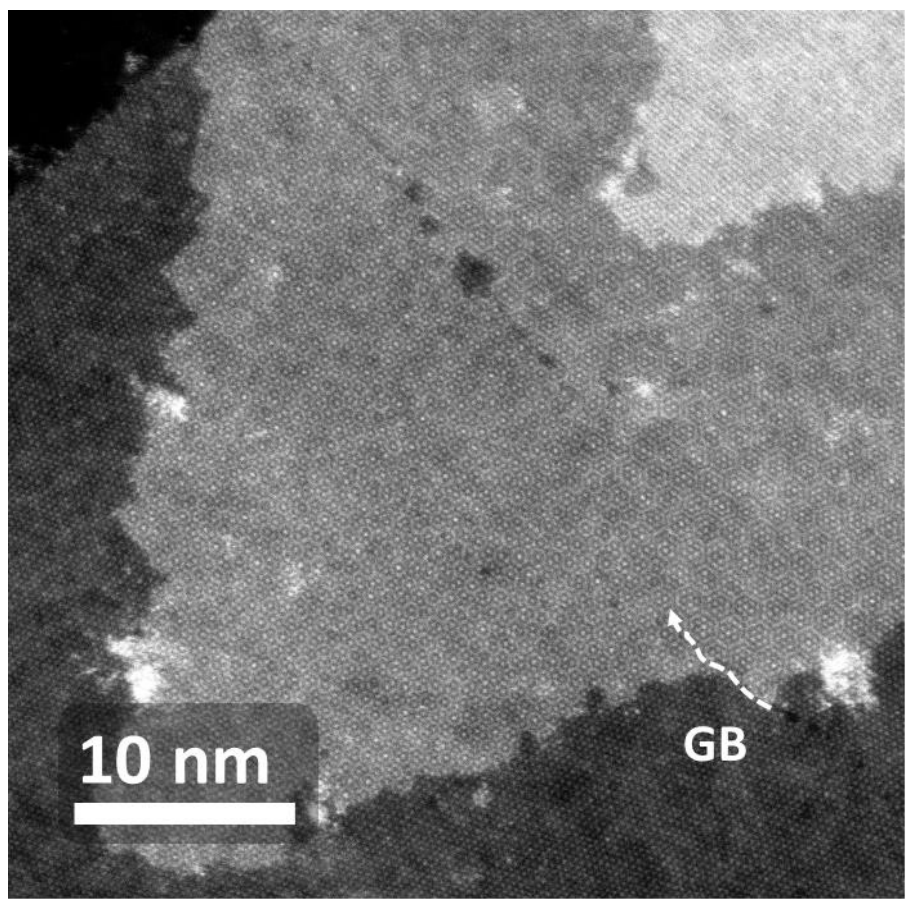

S9: HAADF image from a different region showing the modulation of the moiré pattern due to growth of the 2$^{nd}$ WS$_2$ layer across the grain boundary in the underlying WS$_2$ layer. Here the grain boundary is unambiguously formed in the lower WS$_2$ layer.

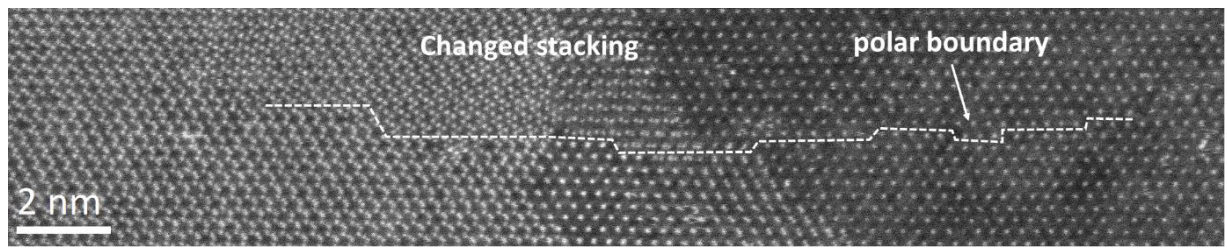

S10: High-resolution image of a polar boundary in the WS$_2$ layer. Similar to the modulation of the moiré pattern across the grain boundaries, here the stacking of the WS$_2$ layers gets modified. Controlling the formation of these polar boundaries could lead to a controllable growth parameter for specific kind of stackings of the WS$_2$.

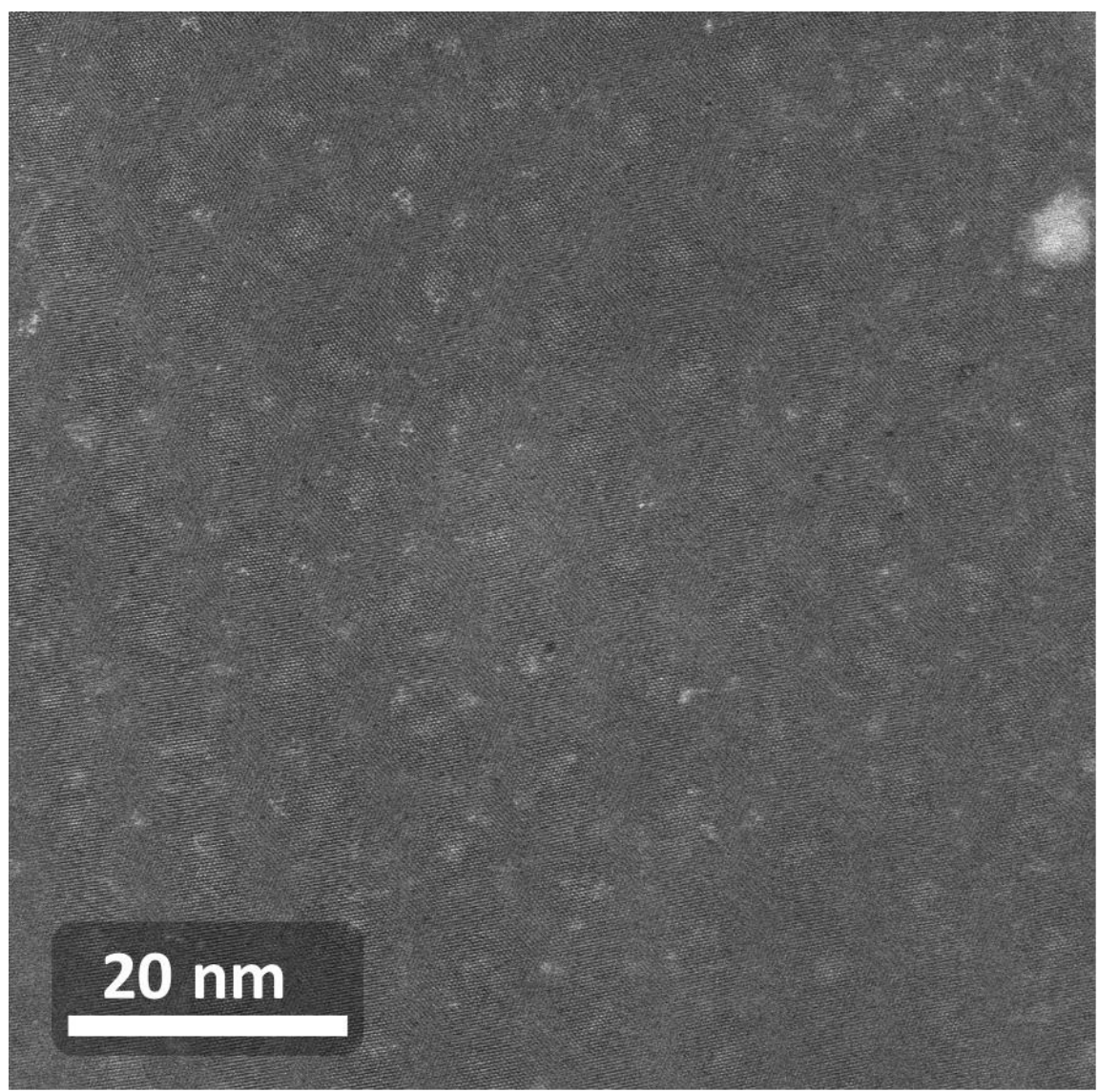

S11: High resolution HAADF image collected at the 2$^{nd}$ moiré area of Figure 5 a) from the main text. The extension of the moiré pattern over several 10$^{th}$ of nm is seen. Furthermore, modulation of the moiré pattern across this area is seen, which coincides to the intensity modulation seen in the main text and Figure S11.

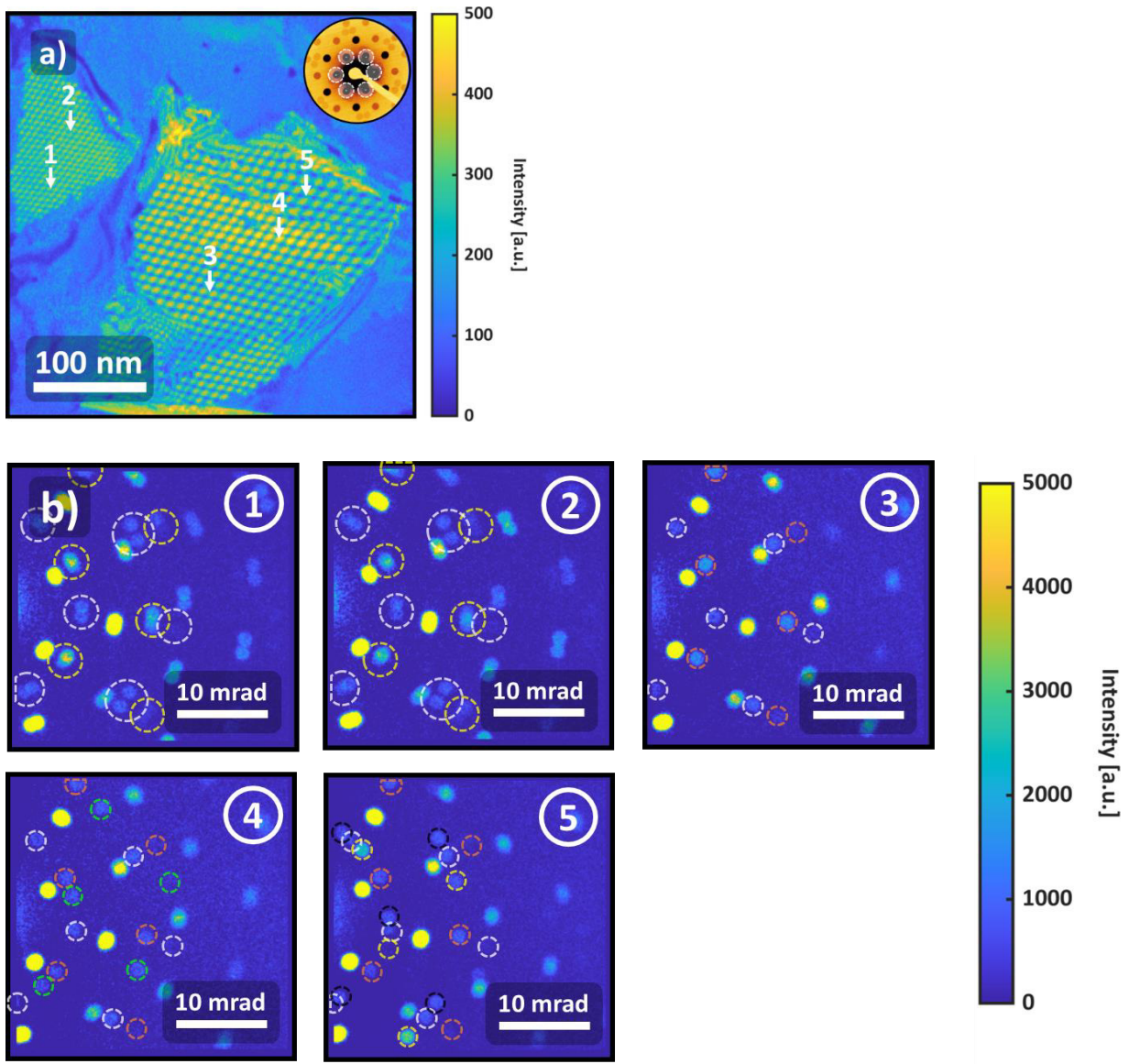

S12: a) Virtual dark field image from Figure 5 a) from the main text showing a region containing 2 low angle $WS_2$ moiré patterns. b) Selected averaged diffraction patterns of 16x16 px from regions 1 to 5, respectively. The colored circles mark the different orientations of the graphene layers.

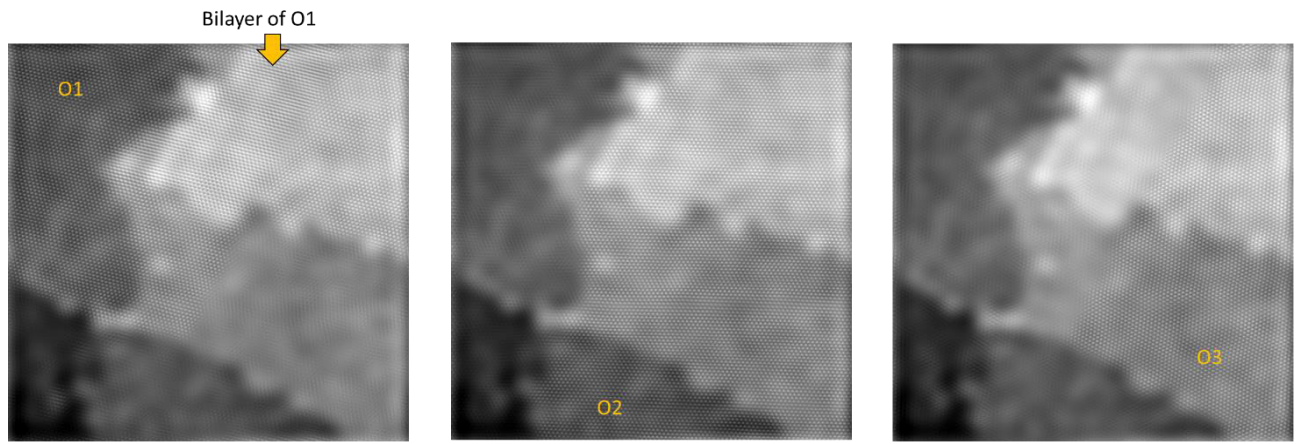

S13: FFT filtered images generated from the high resolution HAADF image shown in Figure 4 a) in the main text. These show the areas in which the 3 different orientations O1-O3 are present in this image.

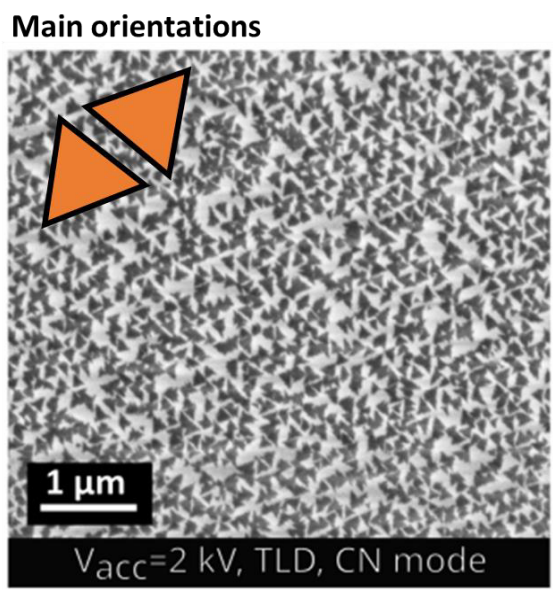

S14: SEM image of $WS_2$ grown for 1800s on sapphire without the FLG layers. Two main orientations as indicated by the orange triangles are observed with respect to the c-plane sapphire. This shows the more stringent alignment of the $WS_2$ with respect to the sapphire, compared to the weakened distribution by the addition of the FLG as seen in Figure 2 b) of the main text.

**References**

[1] Izhak Bucher. Circle fit. Available at https://www.mathworks.com/matlabcentral/fileexchange/5557-circle-fit (2023).

[2] Kazmierczak, N. P. *et al.* Strain fields in twisted bilayer graphene. *Nature materials* **20**, 956–963; 10.1038/s41563-021-00973-w (2021).